# Two-Stage Boost Inverter for Wave Energy Conversion

Souvik Datta[1][0000-0002-8669-3180] and P. Sriramalakshmi[2][0000-0002-8694-4605]

[1,2]School of Electrical Engineering
[1,2] Vellore Institute of Technology, Chennai-600127, India
[1]souvik.datta2019@vitstudent.ac.in
[2]sriramalakshmi.p@vit.ac.in

**Abstract** Mitigation of fossil fuel dependency is becoming a major concern in most of the countries around the globe. In that context, this research article presents the single phase two stage boost inverter system for wave energy conversion. The proposed wave energy based power conditioning system includes wave energy based functional block, permanent magnet synchronous generator, zeta converter, single phase H-bridge inverter and a resistive load. The boost inverter is used to boost up the voltage availed from the permanent magnet generator powered by a wave energy converter system using zeta converter and then invert it using an H-bridge inverter. Hence variable magnitude AC voltage is converted to constant magnitude AC voltage at constant frequency. The blocks of each component is designed to deliver the power of 1kW at the load terminals and simulation study is carried out in the MATLAB / Simulink platform for further analysis. The simulation results are presented and discussed to show how wave powered boost inverter can efficiently convert the wave energy to electrical energy and condition the electrical power before feeding it to any AC load or AC grid. The performance analysis of wave energy conversion system based on zeta converter is included at the end of the article.

## 1 Introduction

### 1.1 Background

The modern world is much concerned about the excessive usage of conventional fossil fuel power plants. These fossil fuel power plants also cause greenhouse effect which is a prime reason for the climate change. Moreover, increasing power demand and the fast depletion of fossil fuels encourage the countries to find alternate solutions such as wind power, photo voltaic and ocean energy. Renewable energy sources will clearly play a vital role in reducing greenhouse gas emissions and ensuring a sustainable future for the entire globe. Development of renewable energy sources, including ocean, tidal and wave energy arises to reduce greenhouse effect and hence improves the production of electricity as well as reduces emission of carbon footprint. To address the aforementioned issues, wave energy has become more popular and various wave energy conversion systems are developed [1][2]. Recently, ocean wave energy is emerging research area since it is considered as one of the cleanest and safest energy sectors. Hence it is gaining more attention among the researchers. Ocean waves are generated by numerous techniques such as through gravity, seismic tremors, solar energy etc. The wave energy potential is calculated theoretically as 16,000 TWh/year [3,4], which is very much helpful in meeting the global energy demand. The wave energy resource with other renewable energy sources, such as wind and solar, can lead to positive synergies [5–7]. Wave energy systems can generate the power up to 90% power from the source, but wind and solar power systems can generate only 20–30% [5–7]. Around 2% of the total energy demand of the globe is met by the ocean wave energy technology [8]. This is one of the cheapest among all other renewable resources, which is mostly rely on photovoltaic and wind power. A comprehensive review on the wave to wire model is presented in [6, 9, 10, 11] which reviews the various methodologies adopted for the conversion process of ocean wave energy to electrical energy using wave to wire model. A series of power electronic converters are essential to convert wave energy to electrical energy before feeding it to the load or grid. The detailed review of wave energy converter topologies is elaborated in [12]. Various electrical generators such as linear and rotary types used in the conversion process are elaborated in [13]. A direct-drive linear generator or rotary generators are used to convert the mechanical energy to electrical energy. The linear type system makes the overall system very economical. Different types of linear generators are available for converting the wave energy into electricity [13]. Based on the geometric and magnetic structures, various configurations of generators like linear and rotational types are available. Among all, the linear tubular generators are found to be highly efficient. The translator may be inside or outside of the generator stator. The power density is not similar for various generators. The linear generators produce the linear motion while the rotational one produces the rotary motion. There are various principles such as Oscillating Water Columns (OWCs), Wave-Activated Bodies (WABs) and Overtopping Devices (ODs), are associated with the wave energy conversion (WEC) technologies. Buoy Type WECs consist of various floating buoy. Very common type of WEC is the Heaving Point Absorber and it takes energy from any direction.

There are various power converter topologies used to convert the wave power to electrical power. There are many WECS technologies available in the literature. But the most promising method of conversion is still not definite for converting wave power to electrical power. The wave energy conversion techniques need to be chosen based on the location to achieve the highest efficiency of conversion. In addition, modeling and control techniques take main role in improving the efficiency of the WECS. Various configurations of wave energy conversion system (WECS) are presented in the literature and each one has adopted different converter topology. The power conditioning topologies are mainly used to adjust and condition the source voltage of the system to attain the desired output voltage. An appropriate combination of a diode rectifier and DC- DC boost converter is used as an interface between the PMSG and a single phase or three phase inverters [4].

In this research article, a wave energy powered zeta converter based two stage boost inverter is used to boost and invert the voltage availed from the wave energy powered rectified dc voltage. AC voltage is generated by a PMSG generator combined with the wave energy conversion system. . The conventional boost converter cascaded with H bridge inverter is already available in the literature. In this article zeta converter based inversion is performed to obtain the constant magnitude of AC voltage from the variable AC voltage generated at the PMS generator. The grid integration aspects are well explained in [15]. A two-stage conversion system to convert variable AC voltage to a constant amplitude AC voltage with constant frequency is implemented using cascaded buck boost inverter system [16]. The complete article is arranged as follows: in section II, the proposed wave energy powered zeta converter-based boost inversion, and the design specification of wave conversion model, PMSG generator design specifications, zeta converter design specifications are tabulated. Section III, discusses the operation of the zeta converter and modulation technique adopted to generate the pulses for the inverter. The simulation results of the completed wave energy based boost inverter are discussed in section IV. The conclusions of the article and the future direction of research in the field of wave energy conversion is presented at the end of the article.

## 2    Circuit Analysis

### 2.1    Wave Energy Powered Boost Inverter System

The block diagram of the wave energy powered zeta converter based boost inverter is shown in Fig.1. The wave power converted using the wave energy converter system is given to the permanent magnet synchronous generator as shown in Fig.1. The electrical energy obtained from the PMS generator is rectified using the AC-DC rectifier and then the converted DC is given as the input to the zeta converter to boost the rectified DC voltage. Further the boosted DC if fed to the single-phase H-bridge inverter for inversion operation. The inverter output is supplied to the resistive load.

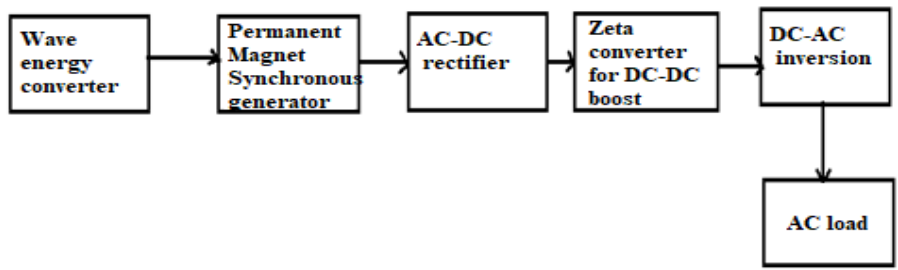

**Fig. 1.** Block diagram of wave energy powered boost inverter

## 2.2 Zeta Converter to Boost the PMSG Output

The zeta converter used to boost the availed rectified DC voltage is depicted in Fig.2. Zeta converter is similar as buck-boost converter. In addition, it has a wider range of duty ratio compared to any other conventional converters. Moreover, Zeta converter has low input current distortion and hence it has an improved power factor. In addition, the output current has lower ripple and output-power range is wide. In this work, the duty ratio of Zeta converter is chosen to act as a boost converter. The boost converter is basically a DC-DC converter in which output voltage is greater than input voltage, while stepping down the current. It has two inductors (L1, L2), two capacitors (C1, C2), switch (TQ1), Diode (TD1). The circuit operation is explained in two modes considering that zeta converter is in Continuous Conduction Mode (CCM) and are presented in Fig.3 and Fig.4 respectively. The frequency of 100 kHz is used to produce the firing pulse for the zeta converter with the duty ratio of 0.697.

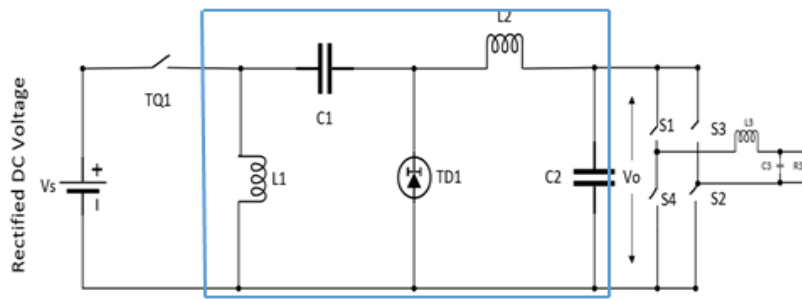

**Fig. 2.** Zeta Converter based boost inverter topology

The topological structure of zeta converter-based boost inverter is depicted in Fig. 2. The Zeta converter operates in two modes –

*1) Zeta Converter in Mode 1:*

During mode 1 as in Fig.3, the MOSFET switch TQ1 is fired on and the diode T$D1$ is reverse biased. The source charges both the inductors L1 and L2. The inductor current $IL1$ and $IL2$ increases linearly. Also, C1 charges C2.

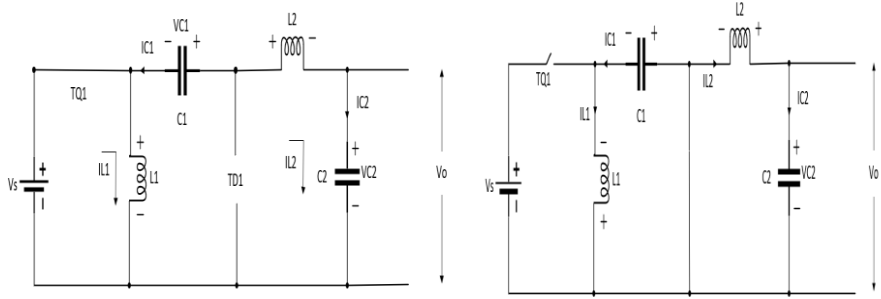

**Fig. 3.** Mode 1 of operation of Zeta Converter    Fig. 4. Mode 2 Operation of Zeta Converter

Inductor voltages ($V_L$) are obtained as –

$$L1 * \frac{dI_{L1}}{dt} = V_s \tag{1}$$

$$\frac{dI_{L2}}{dt} = \frac{V_S}{L2} + \frac{V_{C1}}{L2} - \frac{V_{C2}}{L2} \tag{2}$$

The capacitor charging current is obtained as –

$$C2 * \frac{dV_{C2}}{dt} = I_{c2} \tag{3}$$

*2) Zeta Converter in Mode 2:*

During mode 2 as in Fig.4, the switch $TQ1$ is switched off and the diode $TD1$ is forward biased. Inductors $L1$ and $L2$ discharge through capacitors $C1$ and $C2$. Therefore, the inductor currents IL1 and IL2 decrease gradually.

Inductor voltage ($L_1$) is given as,

$$L_1 \frac{dI_{L1}}{dt} = -V_1 \tag{4}$$

Inductor voltage ($L_2$), is obtained by,

$$L_2 \frac{dI_{L2}}{dt} = -V_{L2} \tag{5}$$

Current through the capacitor $C_1$ is,

$$I_{L1} = C_1 * \frac{dV_{C1}}{dt} \tag{6}$$

$$\frac{V_0}{Vs} = \frac{I_{in}}{I_o} = \frac{D}{1-D} \tag{7}$$

Where D is the duty cycle

$I_{in}$ is the input current through the zeta converter

$V_o$ and $I_o$ are the output voltage across the zeta converter and current through the converter

## 3    Process of converting Boosted DC to AC

The boosted voltage obtained from the zeta converter is fed to H bridge inverter as shown in Fig.1. In this wave powered boost (Zeta) inverter, a resistive load is connected and ac power is delivered to the load. The conventional sinusoidal PWM strategy is used to produce the pulses for the MOSFET switches of the H-bridge inverter. The sinusoidal modulating signal is compared with the high frequency triangular waveform to produce the firing pulses of the inverter.

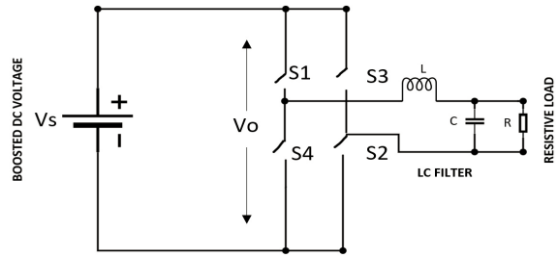

**Fig. 5** H Bridge Inverter

$$V_{acpk} = \frac{MBV_s}{\sqrt{2}}$$

(8)

Where M is the modulation index
B is the boost factor
$V_s$ is the DC output voltage across the rectifier

## 4    Simulation Model of the proposed system

The overall simulation model of the WECS powered boosted inverter system is depicted in Fig.6 [17]. The wave energy conversion is modeled using the functional block with the parameters given in Table.1 and the specifications list of PMSG is listed in Table 2.

**Table 1.** Wave Energy Conversion System Parameters

| Attributes | Values |
|---|---|
| Wave height (H) | 1m |
| Wave period (T) | 10 s |
| Tidal current speed factor ($K_{gc}$) | 1.5 |
| Wave angular frequency (ω) | 2 *pi/T |
| Wave number | 0.408 |

**Table 2.** Specifications of Permanent Magnet Synchronous Generator

| Attributes | Values |
|---|---|
| Number of Phases | 3 |
| Back Emf Waveform | Sinusoidal |

| Rotor Type | Round |
|---|---|
| Stator Phase Resistance ($R_{ph}$) | 0.0484 Ω |
| Armature Inductance ($L_a$) | $3.95*10^{-4}$ H |
| Flux Linkage (ʎ) | 0.1194 |
| Pole Pairs | 8 |

The proposed system is simulated using the specifications of Zeta converter [14] and H-bridge inverter as shown in Table 3 and Table 4.

**Table 3.** Specifications of the Zeta Converter

| Attributes | Values |
|---|---|
| Input DC Voltage (Vs) | 30 V |
| Inductors (L1 & L2) | 1.6 mH |
| Capacitance C1 | 0.159 μF |
| Capacitance C2 | $4e^{-4}$ F |
| Duty Ratio (D) | 0.697 |
| Switching Frequency (fs) | 100k Hz |
| Output Voltage (Vo) | 120 V |

The switching pulse for the zeta converter is produced with the duty ratio (D) of 69.7 % and at the switching frequency of 100k Hz.

**Table 4.** Specifications of H-Bridge Inverter

| Attributes | Values |
|---|---|
| Boosted DC Voltage | 120 V |
| Modulation Index | 1 |
| Switching Frequency | 100 kHz |
| Output Frequency | 50 Hz |
| Output Voltage | 118 V |
| Resistive Load | 15 Ω |
| Filter Inductance | 3 mH |
| Filter Capacitance | 20 μF |

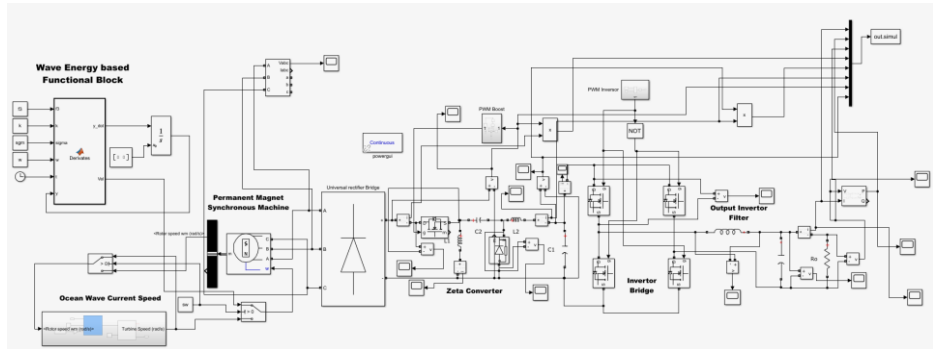

Fig. 6. Simulink Model of the complete wave powered boost inverter system under study

## 5    Simulation Results

The simulated output AC voltage of PMSG is shown in Fig. 7. A three phase AC peak voltage of 30V is obtained from PMS generator. The voltage from PMS generator is given to the diode rectifier for AC to DC conversion.

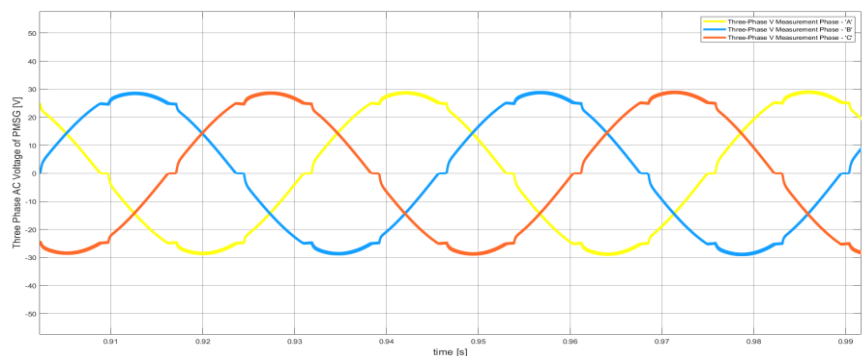

**Fig. 7.** Three Phase AC Voltage of PMSG [V]

A rectified DC voltage of 30V is availed from the rectifier circuit and is depicted in Fig. 8 and is given to Zeta converter for further boosting action.

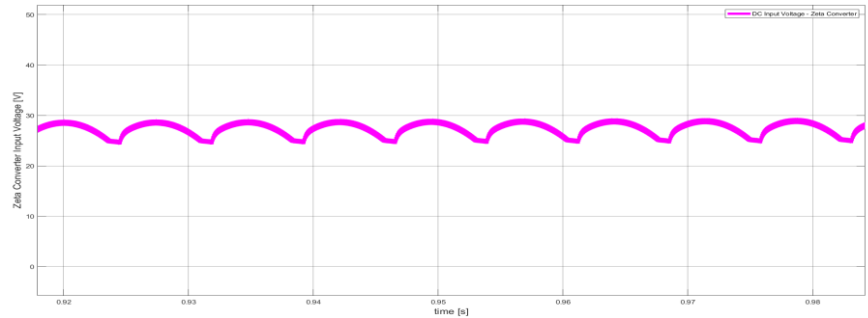

**Fig. 8.** Rectifier output (Zeta converter Input Voltage)

The boosted voltage obtained from Zeta converter is shown in Fig. 9. It is boosted to 120 V DC.

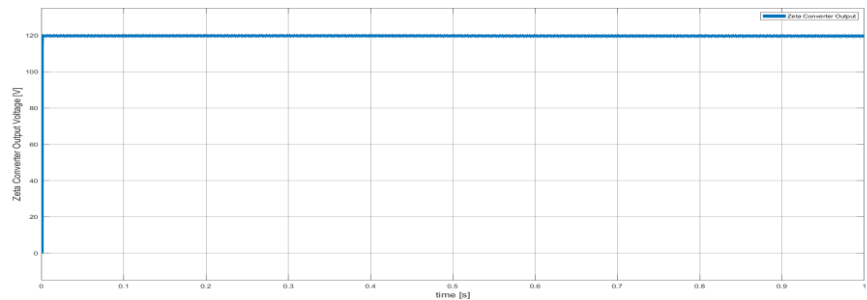

**Fig. 9.** Zeta Converter Output Voltage [V]

The voltage stress across the diode (TD1) which presents in zeta converter is shown in Fig.10. The current flows through the inductors L2 and L1 are shown in Fig.11 and Fig.12 respectively.

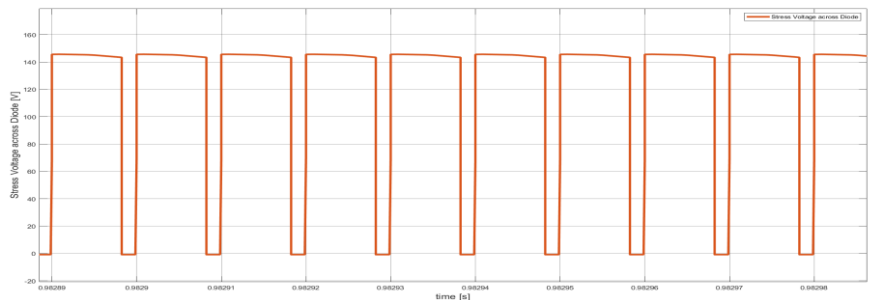

**Fig. 10.** Voltage Stress across Diode [V]

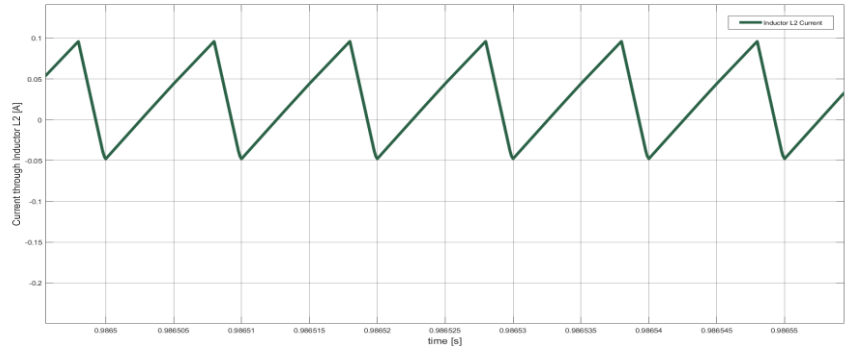

**Fig. 11.** Current through Inductor L2 [A]

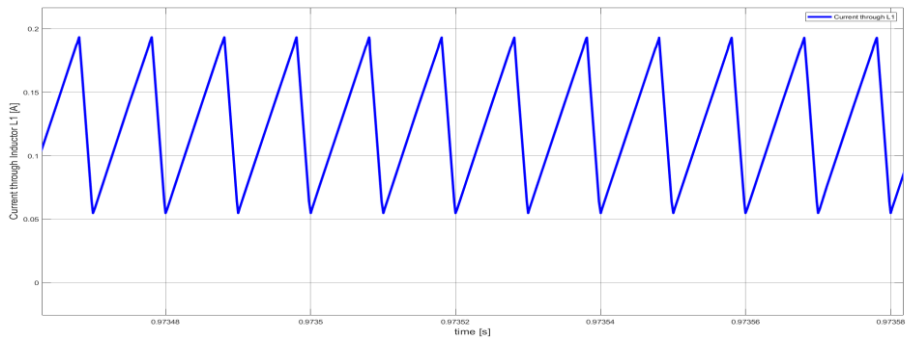

**Fig. 12.** Current through Inductor L1 [A]

The voltage stress across the switch present in the zeta converter is shown in Fig. 13. It is similar to the voltage which is boosted at the zeta converter terminals. The same boosted voltage acts as the input voltage for the H bridge inverter.

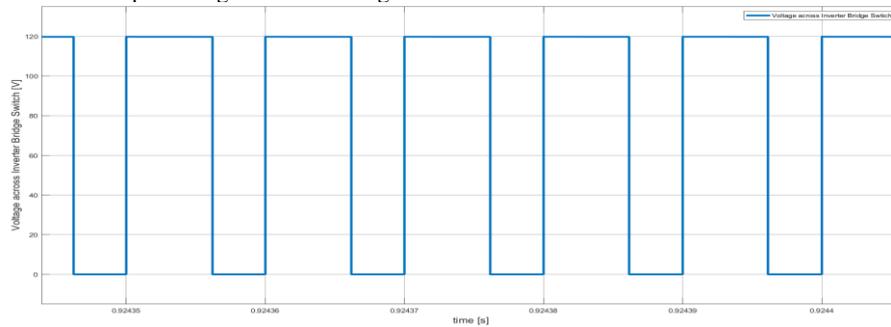

**Fig. 13.** Voltage across zeta converter Switch TQ1 [V]

Fig.14 and Fig .15 depicts the output voltage and output current at the inverter terminals before filter.

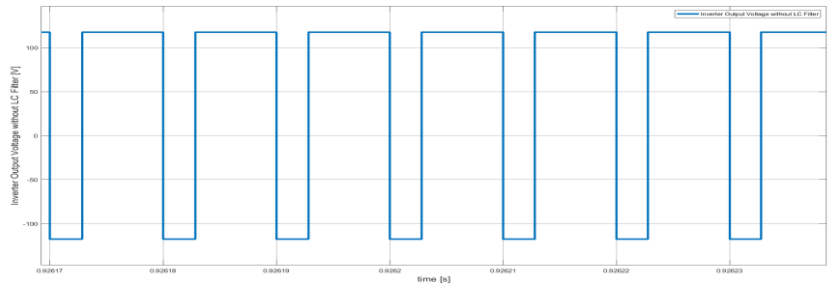

**Fig. 14.** Inverter output voltage without LC filter

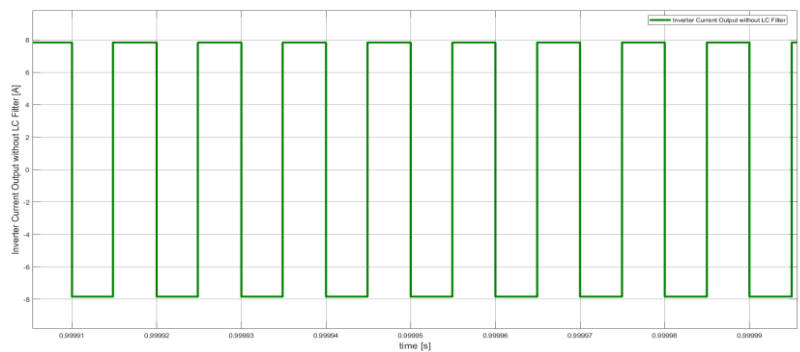

**Fig. 15.** Inverter output current without LC filter

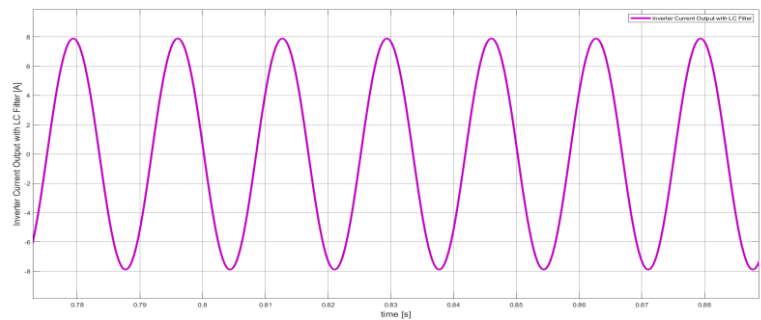

**Fig. 16.** Inverter output current with LC filter

The voltage stress across the switch S1 in the inverter bridge is shown in Fig. 17. The filtered AC inverted voltage obtained from the H Bridge inverter is given in Fig. 18. The filtered peak AC output voltage of 118 V is obtained across the resistive load terminals.

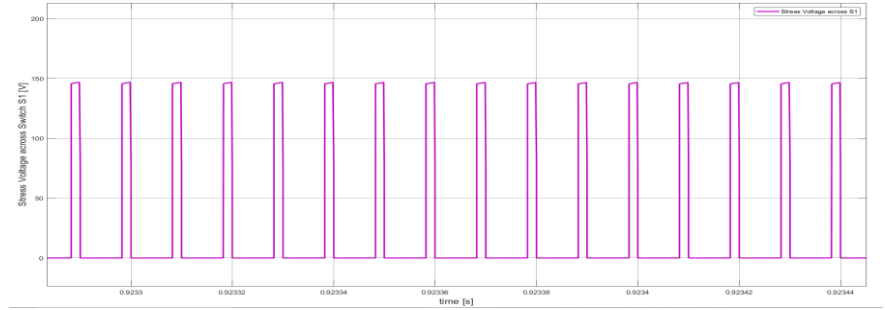

**Fig. 17.** Voltage Stress across Switch S1 [V]

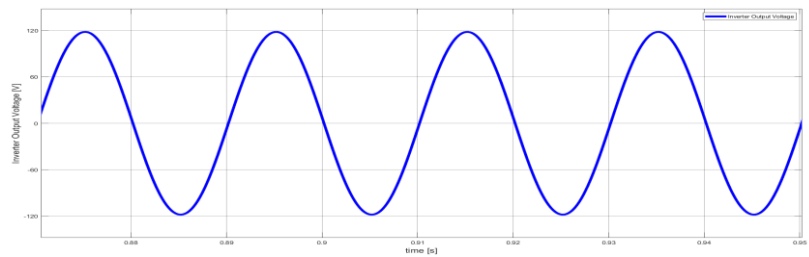

**Fig. 18.** Inverter AC Output Waveform [V]

Fig. 19 (a) shows the relation between duty ratio and the voltage boost occurs in the circuit. For the same duty ratio, zeta converter can provide the higher boost compared to conventional boost converter.

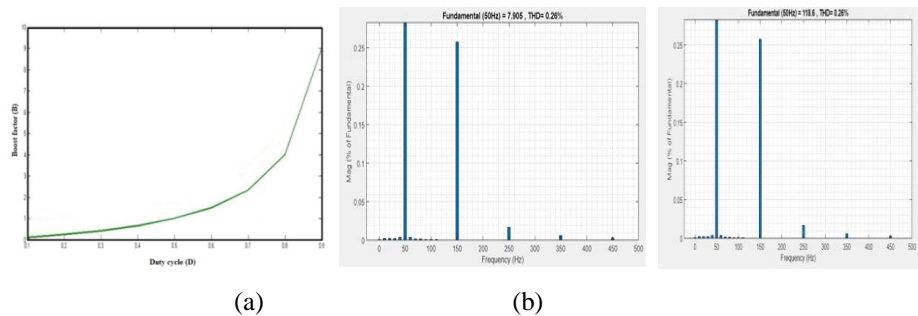

(a)                                        (b)

**Fig. 19** (a) Graph between duty ratio and boost factor   (b) Harmonic profile of load current

(c) Harmonic profile of load voltage

By changing the value of duty ratio of the zeta converter, the conduction period of the switches can be controlled. With the rectified dc voltage of 30 V, zeta converter boosted the voltage to 120 V and with the modulation index of 1, the inverted ac voltage of 118 V is obtained at the efficiency of 98.33% keeping all the devices as ideal during the simulation. Fig. 19(b) and 19(c) show the harmonic profile of the output load cur-rent flowing through the load and voltage across the resistive load at fundamental frequency. The THD content meets the IEEE standards.

# 6    Conclusion

Wave energy technologies are going to be the better replacement and meet the shortage of fossil fuel in the near future. The wave energy is converted into electric energy using modern WECS technologies; but, they are not yet commercialized. And ocean technologies have great potential, but standard policies must be framed to support the research. Efficient WECS need to be developed in the upcoming years. More converter topologies and control strategies need to be developed for efficient conversion. Instead of two stage conversion, single stage boost conversion can be adopted to improve more boost and effective conversion.